\documentclass[preprint,journal]{vgtc}       

\ifpdf
  \pdfoutput=1\relax                   
  \usepackage{graphicx}                
  \DeclareGraphicsExtensions{.pdf,.png,.jpg,.jpeg} 
\else
  \usepackage{graphicx}                
  \DeclareGraphicsExtensions{.eps}     
\fi%

\graphicspath{{figures/}{pictures/}{images/}{./}} 

\usepackage{microtype}                 
\PassOptionsToPackage{warn}{textcomp}  
\usepackage{textcomp}                  
\usepackage{mathptmx}                  
\usepackage{times}                     
\usepackage{cite}                      
\usepackage{tabu}                      
\usepackage{booktabs}                  
\usepackage{makecell}
\usepackage{multirow}
\usepackage{balance}       
\usepackage{paralist}
\usepackage{setspace}
\usepackage{balance}
\newcommand{\DeleteButton}{\includegraphics[scale=0.05]{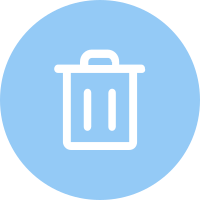}}
\newcommand{\AddButton}{\includegraphics[scale=0.05]{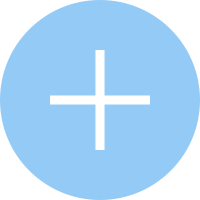}}
\newcommand{\RefreshButton}{\includegraphics[scale=0.05]{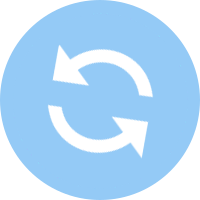}}
\newcommand{\SearchButton}{\includegraphics[scale=0.05]{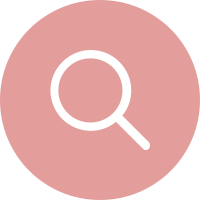}}
\newcommand{\EditButton}{\includegraphics[scale=0.05]{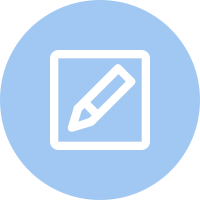}}

\onlineid{1421}

\vgtccategory{Research}
\vgtcpapertype{application/design study}

\title{PromotionLens: Inspecting Promotion Strategies of Online E-commerce via Visual Analytics}

\author{Chenyang Zhang, Xiyuan Wang, Chuyi Zhao, Yijing Ren, \\Tianyu Zhang, Zhenhui Peng, Xiaomeng Fan, Xiaojuan Ma, and Quan Li}
\authorfooter{
\item
 C. Zhang, X. Wang, C. Zhao, Y. Ren and Q. Li are with School of Information Science and Technology, ShanghaiTech University. Quan Li is the corresponding author. E-mail: zhangchy1,wangxy7,zhaochy1,renyj,liquan@shanghaitech.edu.cn.
\item
 T. Zhang is with Geek+. E-mail: macwish@hotmail.com.
\item
 Z. Peng is with School of Artificial Intelligence, Sun Yat-sen University. E-mail: pengzhh29@mail.sysu.edu.cn.
\item
 X. Fan is with School of Entrepreneurship and Management, ShanghaiTech University. E-mail: fanxm@shanghaitech.edu.cn.
\item
 X. Ma is with the Hong Kong University of Science and Technology. E-mail: mxj@cse.ust.hk.
}

\shortauthortitle{Biv \MakeLowercase{\textit{et al.}}: Global Illumination for Fun and Profit}

\abstract{
Promotions are commonly used by e-commerce merchants to boost sales. The efficacy of different promotion strategies can help sellers adapt their offering to customer demand in order to survive and thrive. Current approaches to designing promotion strategies are either based on econometrics, which may not scale to large amounts of sales data, or are spontaneous and provide little explanation of sales volume. Moreover, accurately measuring the effects of promotion designs and making bootstrappable adjustments accordingly remains a challenge due to the incompleteness and complexity of the information describing promotion strategies and their market environments. We present \textit{PromotionLens}, a visual analytics system for exploring, comparing, and modeling the impact of various promotion strategies. Our approach combines representative multivariant time-series forecasting models and well-designed visualizations to demonstrate and explain the impact of sales and promotional factors, and to support ``what-if'' analysis of promotions. Two case studies, expert feedback, and a qualitative user study demonstrate the efficacy of \textit{PromotionLens}.
} 


\CCScatlist{ 
 \CCScat{K.6.1}{Management of Computing and Information Systems}%
{Project and People Management}{Life Cycle};
 \CCScat{K.7.m}{The Computing Profession}{Miscellaneous}{Ethics}
}

\teaser{
  \centering
  \includegraphics[width=\textwidth]{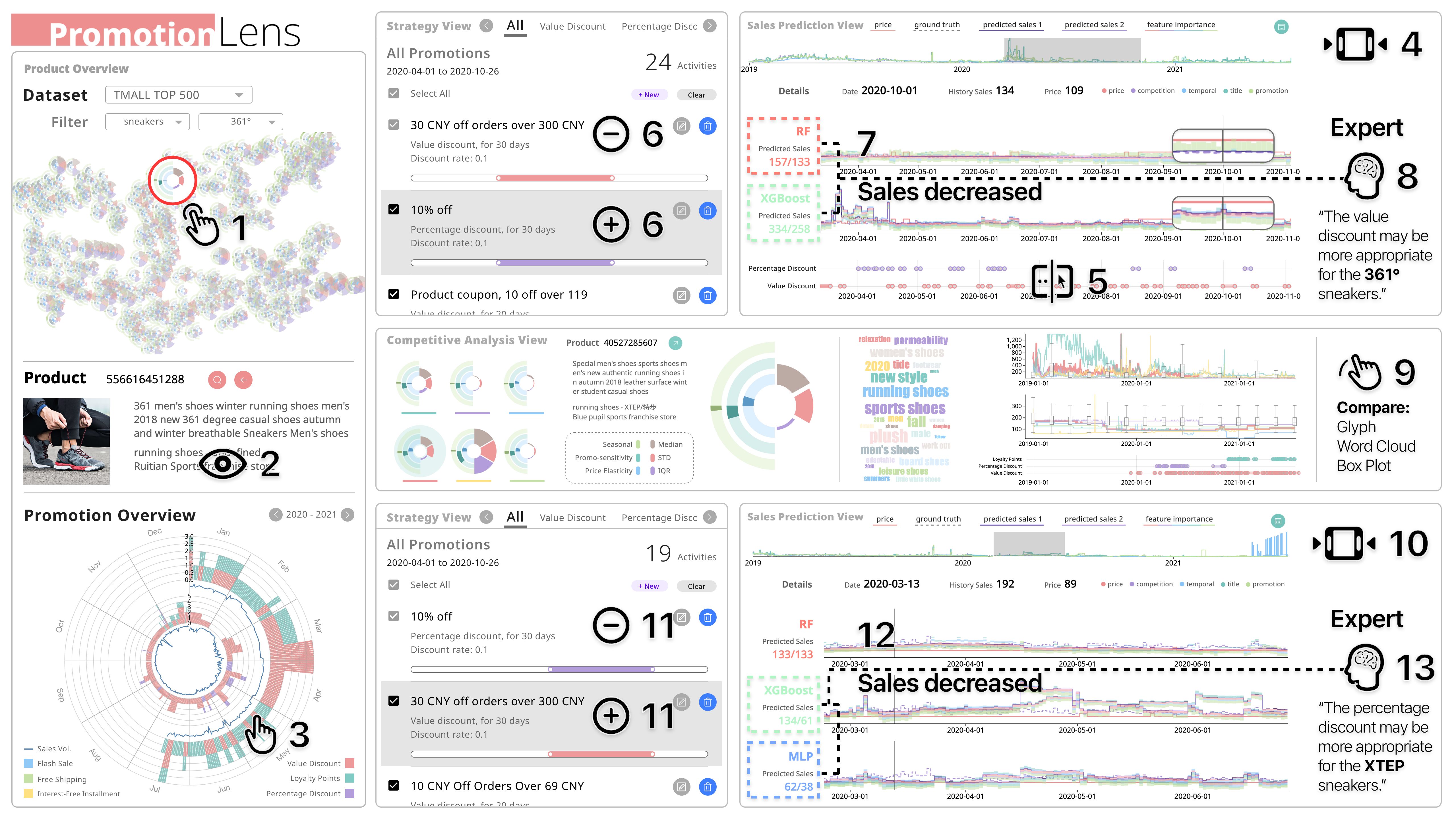}
  \vspace{-8mm}
\caption{Case I: (1) Select a product within the red circle among all 361° sneakers in the \textit{Product Overview}.  (2) Browse the product description of the selected product at the buttom of the scatter chart. (3) View the annual data of the selected product in the \textit{Promotion Overview}. (4) Adjust the time period selector in the \textit{Sales Prediction View}. (5) Look through the dotted lines and find non-overlapping zones. (6) Change the promotion strategy in the \textit{Strategy View} -- exchange the value discount that exists with the same discount rate and percentage discount for the time period. (7) Discover that sales have decreased. (8) Conclude that the value discounts may be appropriate for the selected product. (9) Compare similar products in the glyphs, word clouds and box plots in the \textit{Competitive Analysis View}, and select another XTEP product. (10) Adjust the time selector to fix a time period for the newly selected product. (11) Replace the existing percentage discount in the \textit{Strategy View} with the same discount rate and value discount for the time period. (12) Find that sales have decreased. (13) Conclude that the percentage discount may be appropriate for the selected product.}
  \label{fig:case_1}
}

\vgtcinsertpkg

\begin{document}



\firstsection{Introduction}

\maketitle
\par E-commerce has boomed over the past few decades as more and more transactions can be completed online. The boom has brought competition to e-retailers, who must constantly gain new ideas in promotions to retain and grow customers. Previous studies have shown that promotions in the Internet environment are likely to encourage people to buy goods~\cite{chen1998effects,currim1991taxonomy,gedenk1999role,gedenk2010sales,yusuf2020effect}. Sales rocketed over the periods of successful e-commerce promotions, such as \textit{Black Friday}, \textit{Cyber Monday}, or \textit{Singles Day}~\cite{li2020sustainability,song2020characteristics,xu2020did}. However, simply adding discounts to merchandise is not always desirable, and different promotion strategies and combinations can significantly affect or even change the results. Research on giveaways suggests that if customers pay more than they expect, the corresponding promotion strategy may backfire and reduce sales~\cite{cai2018there}. Similarly, some researchers have noted that even though both promotions can save buyers the same amount of money, there is still a preference for promotional terms between categories~\cite{mishra2011influence}, leading to different performance in sales statistics.

\par Some previous studies have provided reasonable and explainable predictions for sales volume by comparing different machine learning (ML) and deep learning (DL) models to visually dig into the reasons behind each rise in sales~\cite{khakpour2021visual,shao2021visual,sun2020dfseer,xu2021mtseer}. However, to the best of our knowledge, most of these works have focused on the statistical characteristics of time series data rather than incorporating the promotion strategies that most e-commerce retailers must consider in order to gain more attention and expect higher profits. On the other hand, researchers in management and economics have proposed a number of qualitative and quantitative theories for various effects induced by promotions, such as the frequency of hosting discount events~\cite{7874112}, the art of recommendation using complementarity~\cite{wan2018representing}, and the substitution effects~\cite{jiang2015redesigning}. However, few of them are able to scale to larger data volumes, support interactive guidance of promotions, and incorporate predictive capabilities~\cite{chong2016predicting}. Notably, we considered the interests of both parties and tried to answer the following questions: \textit{1) Which promotion strategies are more effective and what are their features?} \textit{2) What would happen to sales volume if other promotion strategies were chosen instead of traditional ones?} \textit{3) What combinations could bring the highest sales?}

\par Nonetheless, examining promotion strategies and analyzing their correlation with sales volume is not straightforward. First, future trends in sales volume can be influenced by a variety of factors, such as the inherent attributes and time-series characteristics of the merchandise~\cite{sun2020dfseer,xu2021mtseer}. However, the analysis of the discount impact would not be convincing if we could not first de-correlate different factors. Second, the diversity of promotion strategies, e.g., \textit{``event sponsoring''} and \textit{``offering refunds''}, also makes it more difficult to accurately understand and predict sales, and therefore requires a great deal of effort to properly quantify and normalize different kinds of promotion strategies. Third, e-merchants usually decide their promotional policies based on their past sales statistics and business experience with specific items. They, especially small and medium-sized e-merchants rarely have the opportunity to quantitatively review the promotion strategies of other merchants. In addition, the risk of betting once on a new promotion strategy prevents them from trying other strategies or combinations that might produce better results than the current situation. Little research has been done on the ``what-if'' analysis~\cite{golfarelli2006designing} of promotion strategies.

\par In this study, we examine promotion strategies in online e-commerce by considering time-series sales data and promotional factors. Notably, we first conduct an observational study of the current practices of our collaborating experts to identify their main needs and concerns regarding the analysis of promotion strategies for e-commerce platforms. We then streamline the analysis of promotions by 
leveraging multiple interpretable machine learning models to identify key factors that have a significant impact on sales volume. We support domain experts in comparing the performance of different products in terms of sales volumes, products descriptions, promotion activities, and related features. We further provide an interactive bootstrap mechanism of promotional features to determine the response of machine learning models. We further propose a visual analytics system, \textit{PromotionLens}, to help end-users select, compare, and combine different promotion strategies. Specifically, it helps users to select target items, check out key promotional factors, and make promotion choices through ``what-if'' analysis. Two case studies, interviews with domain experts, and a qualitative user study validate the effectiveness of our approach. We summarize the main contributions as follows:
\begin{compactitem}
\item We consider time-series correlation and quantitative promotion strategy factors for sales volume forecasting.
\item We propose a visual analytics system with new features to support inspection, comparison, and ``what-if'' analysis of multiple adjustments to promotional policies.
\item We conduct two case studies, interviews with marketing researchers and e-merchants, and a qualitative user study.
\end{compactitem}

\section{Related Work}
\subsection{Visualizations in E-commerce}
\par Recently, e-commerce has paid increasing attention to big data analysis~\cite{akter2016big}, in which visualization techniques have been used in depth~\cite{10.1145/3308560.3316605,10.1007/3-540-44934-5_8,4680367}. For example, Huang et al.~\cite{huang2021visual} developed a visual analytics system to classify customers into categories in order to explore shopping patterns in each category. \textit{VAET} extracts timestamped events and anomalies in time series to explain the correlations between salient transactions and temporal events~\cite{xie2014vaet}. \textit{DAV} is a system for visually correlating product affinities for large amounts of e-commerce transaction data~\cite{10.1007/978-3-7091-6215-6_20}. Ko et al.~\cite{https://doi.org/10.1111/j.1467-8659.2012.03117.x} developed a system for analyzing competitive advantage using point of sale data. Brainerd et al.~\cite{963293} developed an interactive, scalable visualization tool to analyze user behavior by using clickstream data on e-commerce websites. Kohavi et al.~\cite{kohavi2004lessons} reviewed the architecture and discussed lessons and challenges of mining retail e-commerce data. In this work, we focus on exploring the relationships between various promotion strategies and the sales performance of e-commerce items. To the best of our knowledge, \textit{PromotionLens} is the first visualization system that supports users in quantifying, comparing, and performing ``what-if'' analysis of promotion strategies based on interpretable time-series ML models.

\subsection{promotion Strategy Analysis}
\par Studies focusing on the analysis of the correlation between promotions and sales amounts have been proposed from different perspectives dealing with different complex situations. For example, Agmeka et al.~\cite{AGMEKA2019851} started from the relation of brand reputation and actual payment actions, concluding that marketers who do discount more often will earn themselves a good reputation and higher selling amount. Other scholars focused on the choice of promotion strategies~\cite{JIANG2021102612}. Some have studied the effect of promotion duration~\cite{LI2020106640}, while others have examined whether complementary strategies are better than current strategies across products. There are also studies on discount pricing, which focus on designing optimal discount policies to maximize the firm's profits~\cite{12774,YUE2013492}, mainly seeking optimal discount pricing strategies to attract consumers in order to increase revenue. According to~\cite{LI2020106640}, in the era of e-commerce marketing, various promotional policies are intensively used and consumers are more rational in their online shopping. For example, research on giveaways suggests that if customers pay more than they expect, the corresponding promotion strategy may backfire and reduce sales~\cite{cai2018there}. Similarly, some researchers point out that even though both promotions can save buyers the same amount of money, there are still preferences for promotion terms between different categories of goods~\cite{mishra2011influence}, leading to different performance in sales statistics. Therefore, combining these factors and sales volume has generated new meaningful research questions in e-commerce. However, most traditional approaches usually make conclusions more qualitatively, which is difficult for machines to understand and quantify into the same unit. In order to create a system that can reasonably consider the effects of a wide variety of promotion strategies, we characterize the design requirement in an e-commerce scenario and seek to quantify the discount measures in the dataset into features of the same unit, which is essential to provide a reasonable result.

\subsection{Time-series Forecasting and Evaluation}
\par Anticipating upcoming trends is one of the hottest topics in the e-commerce industry, usually based on time series forecasting with past statistics, and plays a key role in e-commerce decision making. According to the model taxonomy~\cite{agrawal2013introductory}, time series models can be classified into four groups, namely, \textit{auto-regressive models}, \textit{exponential smoothing (ES) models}, \textit{machine learning models}, and \textit{neural network models}. As models evolve, the challenge of receiving information, classifying it, and selecting the appropriate model become increasingly difficult for human analysis. As a result, researchers have started to propose various methods to help users understand the output and compare different predictive machine learning models. For example, DFSeer~\cite{10.1145/3313831.3376866} imports automatic model selection methods in time series models to help users make better prediction choices. However, it still cannot support making forecasting based on multiple variables, which is inconsistent with most cases in practice. Therefore, multivariate time series models have been proposed to adjust for chance and make more realistic forecasts. For example, Lewis et al.~\cite{lewis1985prediction} proposed an autoregressive model fitting method that can give satisfactory results for bivariate ARMA models. Recently, the development of deep neural networks equipped with e.g., recurrent neural networks (\textit{RNNs})~\cite{che2018recurrent}, multi-layer perception (\textit{MLP})~\cite{cankurt2016tourism}, and convolutional neural networks (\textit{CNNs})~\cite{wang2019multiple} has taken the time series prediction to a new level. In this work, we employ some of the most representative machine learning algorithms for future prediction based on historical commodity data. 

\par In addition to prediction, various evaluation methods have been proposed in the machine learning, visualization, and human-computer interaction communities to compare the performance of different time series prediction models. Notably, the most basic approach is to use various metrics such as mean squared error (\textit{MSE})~\cite{armstrong1992error,hyndman2006another}, scale-dependent measures such as root \textit{MSE} (\textit{RMSE})~\cite{camastra2009comparative,khair2017forecasting}, percentage errors such as mean absolute percentage error (\textit{MAPE})~\cite{barrow2010evaluation,sharda1992connectionist}. Although these basic methods can evaluate different multivariate prediction machine learning models, they do not provide detailed explanations about model comparisons, evaluation, and feature importance. Recently, Xu et al.~\cite{xu2021mtseer} proposed \textit{mTSeer}, a visual analytics system to compare and evaluate different multivariate time series forecasting models in a bootstrappable manner. Similar to their work, we use the importance of various features to explain the sale volume forecasts. However, we differ from \textit{mTSeer} in that we further support ``what-if'' analysis, allowing domain experts to fine-tune, e.g., change, combine the features extracted from promotion strategies and learn the response of the model.

\section{Observational Study}
\subsection{Experts' Conventional Practice and Bottleneck}
\par To understand how e-commerce retailers develop and determine promotion strategies, we work with a panel of five domain experts from a partnering retail company and a local university's entrepreneurship and management departments. The group includes a researcher whose expertise involves consumer information processing and awareness (E1, female, age: $34$), an e-commerce platform manager (E2, male, age: $36$), a retail data analyst (E3, male, age: $25$) and two promotion strategy marketers (E4, female, age: $27$ and E5, male, age: $28$). A considerable part of E2 -- E5's work is to develop a reasonable and practical promotion plans before the promotion season (e.g., Singles' Day, $618$, and Black Friday), so as to increase product sales and try to help e-commerce retailers achieve their profit targets. The retail platform manager (E2) said that for small and medium-sized retailers, the discount sale season is one of the best opportunities to boost sales and revenue. The experts shared the typical process most retailers use to develop a promotional plan as follows. First, prior to the discount promotion season, retailers will do a ``pre-promotion'' campaign to increase sales, and prices will be as low as possible. As a result, customers can potentially be absorbed into the promotional merchandise and may help promote the discount campaign among the target consumers. Through this opportunity, retailers can also get to know their target consumers and do more personalized marketing. They will also promote the merchandise in other unofficial channels, such as social media, to increase the retailers' exposure. The next step is to put the discount promotion strategy into action. Merchants will consider current unit prices, measure discount rates and corresponding plans, and estimate the associated sales volume. They also partner with other channels to drive customer flow, e.g., using competitive search rankings provided by the platform to increase exposure.

\par Although domain experts have designed and implemented a number of marketing campaigns and strategies, they all agree that the current promotional programs are based only on past retail experience and crude sales data analysis. There are more factors to consider, and the experts mainly encounter the following problems. First, the experts mentioned that in most cases, they would discount at half price, which is convenient and has the potential to increase sales to some extent. However, on the other hand, promotion strategies like ``half price'' can lead to a sharp drop in margins and destroy profitability. Therefore, the retail team needs a demand forecasting model to help them get a handle on the relationship between sales and prices at any given time, as well as other possible factors, ``\textit{such as seasonality and promotions,}'' said E2. Second, retail experts said that promotional programs should be designed differently depending on the different price elasticities (i.e., the degree of change in market demand caused by price changes of the items sold). For example, for items with zero price elasticity, considerable discounts will not increase sales, but will erode margins. For those goods with high price elasticity, a small discount can greatly increase sales while margins or sales are slightly reduced. Therefore, designing promotion strategies should pay close attention to individual items. Third, identifying a specific promotional plan during the traditional e-commerce promotional season is more like a gamble of opportunity. There is no way to experience other promotional programs or to anticipate their corresponding impact on sales. ``\textit{Promotions mean higher customer flow and sales, and it would be constructive if we could predict ahead by trying other promotion strategies,}'' said E5.

\begin{figure*}[h]
    \centering 
    \includegraphics[width=\textwidth]{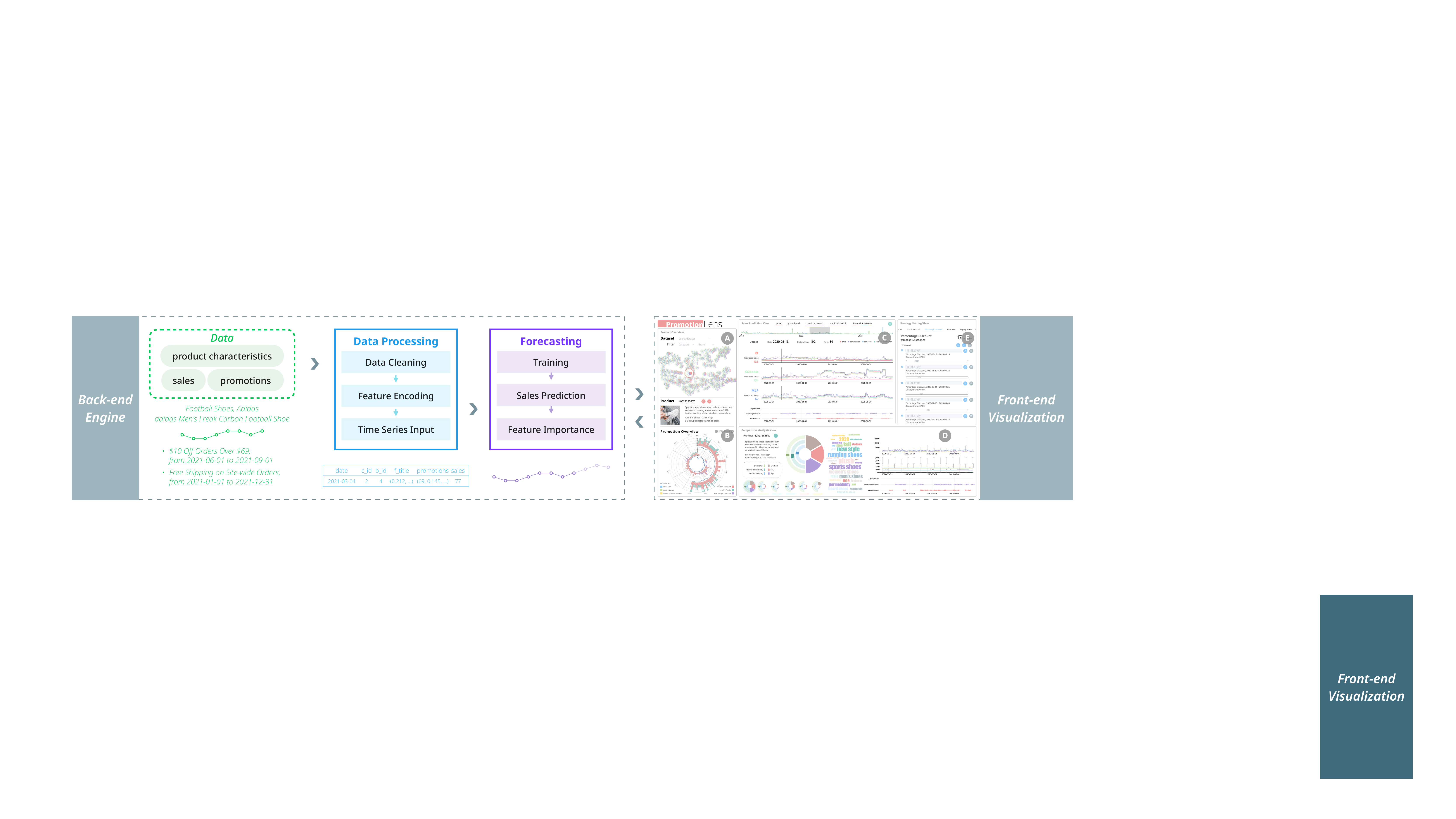}
    \vspace{-6mm}
    \caption{The architecture of \textit{PromotionLens} consists of the back-end engine and front-end visualization. (A) The \textit{Product Overview} helps users select products they are interested in, discover relationships between other products, and get information about their titles, brands, and stores. (B) The \textit{Promotion Overview} shows the development of a product's promotions and how they relate to sales over the last two years. (C) The \textit{Sales Analysis View} shows the forecast of the three models and the duration of the promotions. (D) The \textit{Competitive Analysis View} shows the data of the five most similar competitors for comparison. (E) The \textit{Strategy Setting View} allows to set customized promotion strategies.}
    \label{fig:UI}
    \vspace{-6mm}
\end{figure*}

\subsection{Experts' Needs and Expectations}
\par To ensure that our approach was well suited to the tasks and requirements of the general field, we further interviewed all experts (E1 -- E5) to identify their main concerns about developing promotion strategies. At the end of the interviews, the need for a highly interactive visualization system emerged as a key theme in the feedback collected. While individual expectations for such a system had different emphasis, specific design requirements were expressed across the board.

\par \textbf{R.1 Enable interactive comparison of products.} Marketers often compete with peers selling similar products and need information on other complementary products to make promotional decisions accordingly. Therefore, experts expected that our approach can provide an overview of available product data, including but not limited to \textit{product ID}, \textit{category}, and \textit{brand}, to help them observe relationships with other items and quickly retrieve specific products for further analysis.

\par \textbf{R.2 Predict future trends for individual items.} Traditionally, when estimating and forecasting sales, our domain experts obtain historical sales data from Enterprise Resource Planning (ERP) system and invite some univariate time series models to make forecasts. ``\textit{The system doesn't perform as well as expected, especially during promotional seasons,}'' said E3. ``\textit{We can empirically add a linear factor determined by promotion discount rates, but the results are often unsatisfactory,}'' said E2. E1 also said that multiple variables combine to determine the amount of sales, of which ``\textit{some are difficult to quantify.}'' She also commented that ``\textit{promotion is one of the biggest factors that motivate shoppers to press the `buy button'.}'' Therefore, our approach should support multivariate time series forecasting, consider quantifiable factors, and add promotion features to adjust the prediction results.

\par \textbf{R.3 Understand the factors that influence sales volume throughout the product lifecycle.} According to the feedback from the domain experts, it is important not only to get the best promotion strategy, but also to find out what causes the diversity of results. E2 and E3 were particularly interested in some rare and unexpected situations in the timeline. Therefore, equipping the visualization system with factor analysis allows them to easily observe and speculate on the influencing factors. E1 also mentioned that advanced analytics should get feedback during the sales evolution of the merchandise and that ``\textit{retailer should be aware of influencing factors that may affect the sales volume for the whole sales period and not be limited to a certain promotion phase}.''

\par \textbf{R.4 Identify the similarities and differences of existing promotion strategies.} A promotion is a short-term incentive activity designed to encourage people to buy a product or service. Although there are many dazzling promotions such as \textit``{coupons}'', \textit``{gifts}'', and \textit``{station-wide discounts}'', according to E4 and E5, there are many misconceptions in designing promotion strategies. For example, they said that it is easy for some stores to make promotions regular. ``\textit{Promotions should be short-term. If they become routine, product prices become everyday prices, which translates into no promotions,}'' said E5. ``\textit{Once merchants stop promotions, customers will hesitate and refuse to buy until they are promoted,}'' which can also cause significant damage to a brand's reputation. In addition, many existing promotion strategies are relatively similar, and some retailers simply follow their competitors without thinking about why such promotion strategies work. ``\textit{Some merchants have only one promotional policy, for example, `always discounts',}'' said E4. Therefore, domain experts asked our system to support a detailed exploration of the similarities and differences of common promotion strategies available on e-commerce platforms.

\par \textbf{R.5 Support ``what-if'' promotion simulations for each commodity.} According to E1, there is always a strong need and demand for promotion strategy development. Although previous studies have conducted theoretical and mathematical simulations of specific promotional behaviors, experts still wanted a system that can: enable interactive adjustment of promotional policies, demonstrate the effects of different promotion strategies on a specific commodity, and summarize the effects of promotional behaviors through the performance on various commodities. Thus, our system should support ``what-if'' promotion simulations to examine the model's response for each commodity.

\section{Approach Overview}
\par We propose \textit{PromotionLens}, a visual analytics system for evaluating promotion strategies based on time-series predictive models. \autoref{fig:UI} shows the architecture of \textit{PromotionLens}, which consists of a \textit{back-end prediction engine} composed of a \textit{data processing module} and a \textit{forecasting module}, followed by a \textit{front-end visualization}. To be precise, before building the sales forecasting module, the data processing module quantifies the qualitative text descriptions in the raw data and transforms them into features that are acceptable to the data acceptance port for time series forecasting. After that, various predictive machine learning models are first trained based on the inputs, and then the prediction module is employed for sales prediction and interpretation. At this point, we reach the feature importance of each model. We provide several elaborate visualizations in the subsequent front-end visualization to enable domain experts to make interactive comparisons between products, predictive models, and promotion strategies.

\section{Back-end Engine}
\par In this section, we describe the mechanism of how the back-end engine works, including adapting raw data collected from a large e-commerce platform into quantifiable features that can be used in predictive models and to forecast sales under specific promotion strategies.

\subsection{Data Processing}
\par We used a dataset collected from a e-commerce platform with hundreds of millions of consumers, through a partnership with retail companies, to cover the top $500$ products in the sportswear market in terms of sales volume over a two-and-a-half-year period from January $2019$ to July $2021$. It is worth noting that there are three types of data: 1) \textit{Sales Volume.} The daily sales volume and cumulative sales volume of different products are recorded; (2) \textit{Product Description.} Daily information such as product name, price before discount, category, brand, and merchant to which the product belongs is recorded; (3) \textit{Promotions.} All promotions experienced by each product are recorded, including a detailed description of the event (e.g., \$10 Off Orders Over \$69), the start time, and the end time. Before modeling, several non-numeric dimensions in the original dataset need to be quantified, such as: \textit{category}, \textit{brand}, \textit{product name}, and \textit{promotion details}. Some of these attributes have a limited range of values and are relatively fixed. It is acceptable to simply map them to numbers to maintain information. However, with respect to titles and promotion details, which are a series of descriptive words or semantic sentences, it is expensive to use a one-to-one mapping because these words may be different for different products at different times. Therefore, a well-designed quantification approach to product names and promotional details is needed.

\par Designing appropriate product titles is one of the most fundamental ways to attract potential customers by hitting their searches. The resulting titles are crucial for machine learning models to distinguish between various products. As mentioned earlier, product titles are composed of descriptive words that may not be logically connected to each other and can therefore be treated as a set of words rather than parsed from the perspective of a complete sentence. For this purpose, we utilize a bag-of-words (\textit{BOW})~\cite{zhang2010understanding} method to obtain the information hidden in the titles. Specifically, we first create a codebook of all the words that appear in product titles. Then, we split each title into a set of terms and compare them to the codebook to determine which words appear in the title. In this way, each title can be encoded as a vector with either $0$ or $1$, where $0$ indicates the corresponding keyword in the codebook is missing from the title and $1$ indicates that the keyword is present in the title. It should be noted that the dimensionality of the constructed vector is quite high for a title with only ten words, which is too sparse for the model to learn features from the title. To solve the sparsity problem, we follow the idea of \textit{word2vec}~\cite{church2017word2vec} and reduce the title dimension to $8$ after several experiments, i.e., we calculate the number of semantically valid words in $100$ titles. The results show that eight words are sufficient to describe a product.

\par Although promotions can be diverse, they usually follow some basic rules that allow us to translate textual information into numerical values. Based on these rules and the recommendations of experts, we classify all promotions into two categories, including six promotion types (\autoref{tbl:promotion_types}), for precise quantification. For direct discounts (\textit{value discount}, \textit{percentage discount}, and \textit{flash sale}), we extract two key features. (1) the discount rate $k_d$, which is the ratio of the discounted price to original price; (2) the trigger amount $p_t$, which is the minimum amount to receive the discount. For indirect discounts (i.e., \textit{loyalty points}, \textit{free shipping}, and \textit{interest-free installment}), where the reward is not a direct price reduction, the reward and the original price of the product are taken out as features. In addition, since the effect of a promotion may exist before the campaign starts or after it ends, we track the entire lifecycle of the promotion by setting a status flag.

\begin{table}[h]
\centering
\small
\begin{spacing}{1.2}
\caption{Promotion Types.}
\begin{tabular}{cll}
\toprule
\multicolumn{1}{c}{Category}       & \multicolumn{1}{c}{Type}  & \multicolumn{1}{c}{Form}                    \\
\midrule
\multirow{3}{*}{\makecell[c]{Direct\\ Discount}}   & \makecell[c]{Value Discount}       & \textit{\$10 Off Orders Over \$69}          \\ \cline{2-3} 
                                   & \makecell[c]{Percentage Discount}       & \textit{20\% Off}                           \\ \cline{2-3} 
                                   & \makecell[c]{Flash Sale}                & \textit{30\% Off in 4 Hours}                \\ \hline
\multirow{3}{*}{\makecell[c]{Indirect\\ Discount}} & \makecell[c]{Loyalty Points}                & \textit{100 Loyalty Points Back}            \\ \cline{2-3} 
                                   & \makecell[c]{Free Shipping}                 & \textit{Free Shipping on Orders Over \$99}  \\ \cline{2-3} 
                                   & \makecell[c]{Interest-free Installment}     & \textit{6 Months Interest-free Installment} \\
\bottomrule
\end{tabular}
\label{tbl:promotion_types}
\end{spacing}
\vspace{-6mm}
\end{table}

\subsection{Forecasting}
\par In order to select the most appropriate model(s), we conduct a series of experiments and derive \autoref{tbl:model_performance}. To take full advantage of the historical time series data, as suggested by E.1, we enter the historical average sales into the model, including the average daily sales for the last month, quarter, half year and full year. Considering the performance and the interpretability of the model, we select three representative multivariate time series forecasting models as candidate forecasting techniques (i.e., RandomForest~\cite{ho1995random}, XGBoost~\cite{chen2016xgboost}, and MLP~\cite{hastie2009elements}). Our system can easily integrate other types of time series forecasting models, e.g., Vector Auto-Regressive Model (VAR), RNN, and CNN. Feature importance measures the impact of each feature on the prediction results, and explains the model's decision by calculating the additivity of each feature. We apply SHapley Additive exPlanations (SHAP)~\cite{NIPS2017_7062}, an interpretation method that calculates Shapley values at the instance level, to estimate the feature importance for the sales prediction results. The Shapley values of these features add up to five values (i.e., \textit{descriptions}, \textit{price}, \textit{temporal information}, \textit{competitive information}, and \textit{promotion}), indicating the power from the five relevant aspects. It is worth noting that the Shapley value can be positive or negative, indicating that the feature contributes positively or negatively to the prediction.

\begin{table}[h]
\centering
\small
\vspace{-3mm}
\begin{spacing}{1.2}
\caption{Model performance.}
\begin{tabular}{r@{\hspace{30pt}}c@{\hspace{20pt}}c@{\hspace{20pt}}}
\toprule
{Models}  & {RMSE} & {MAPE} \\ \midrule
Linear           & 4889.36       & 97.21\%       \\ \hline
RandomForest     & \textbf{344.93}        & \textbf{35.21\%}       \\ \hline
XGBoost          & \textbf{322.34}        & 57.69\%       \\ \hline
MLP              & 389.83        & \textbf{31.29}\%      \\ \hline
GradientBoosting & 626.87        & 71.64\%        \\ \bottomrule
\end{tabular}
\label{tbl:model_performance}
\end{spacing}
\vspace{-3mm}
\end{table}

\section{Front-End Visualization}
\par The fundamental design principle of \textit{PromotionLens} is to leverage or enhance familiar visual metaphors so that experts can focus on analysis. We follow the mantra ``\textit{overview first, zoom and filter, then details-on-demand}''~\cite{shneiderman2003eyes}, to guide domain experts to explicitly explore, compare, and model the impact of various promotion strategies. Based on these principles and the previously mentioned requirements, we design and develop the \textit{PromotionLens} interface, which consists of four main views: the \textit{Product Overview} enables domain experts to select products of interest and understand their basic information, including their sales characteristics, product descriptions, and promotion profiles (\textbf{R.1}); the \textit{Sales Prediction View} displays the sales volume predicted by the multivariate forecasting model and lists the promotions in order, along with the influencing factors (\textbf{R.2, R.3}); the \textit{Strategy Setting View} gives a detailed description of the promotion, allowing the domain experts to configure different promotion strategies and simulate their corresponding impact on sales volume (\textbf{R.4, R.5}); the \textit{Competitive Analysis View} shows the characteristics of competing products, including their promotional information, and thus helps to compare promotion strategies and motivate domain experts to find better ones (\textbf{R.1, R.4}).

\begin{figure}[h]
    \centering
    \includegraphics[width=0.44\textwidth]{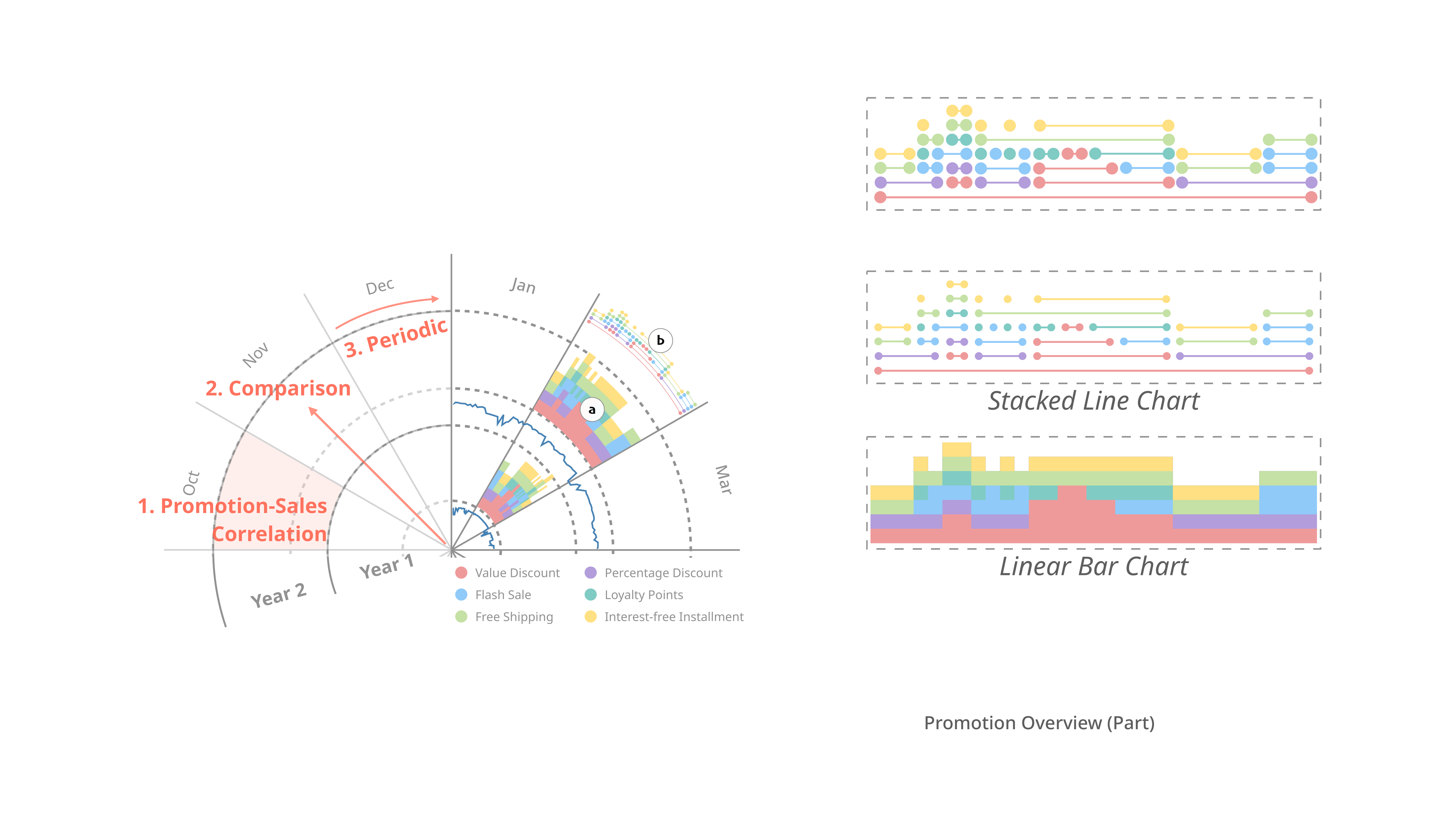}
    \vspace{-4mm}
    \caption{The left part of the figure shows the meaning of each section. The right part shows the details and manner of the promotion (a) and its design alternatives (b): (a) linear bar chart, (b) stacked line chart.}
    \label{fig:promotion_overview}
       \vspace{-6mm}
\end{figure}

\begin{figure*}[h]
    \centering
    \includegraphics[width=\textwidth]{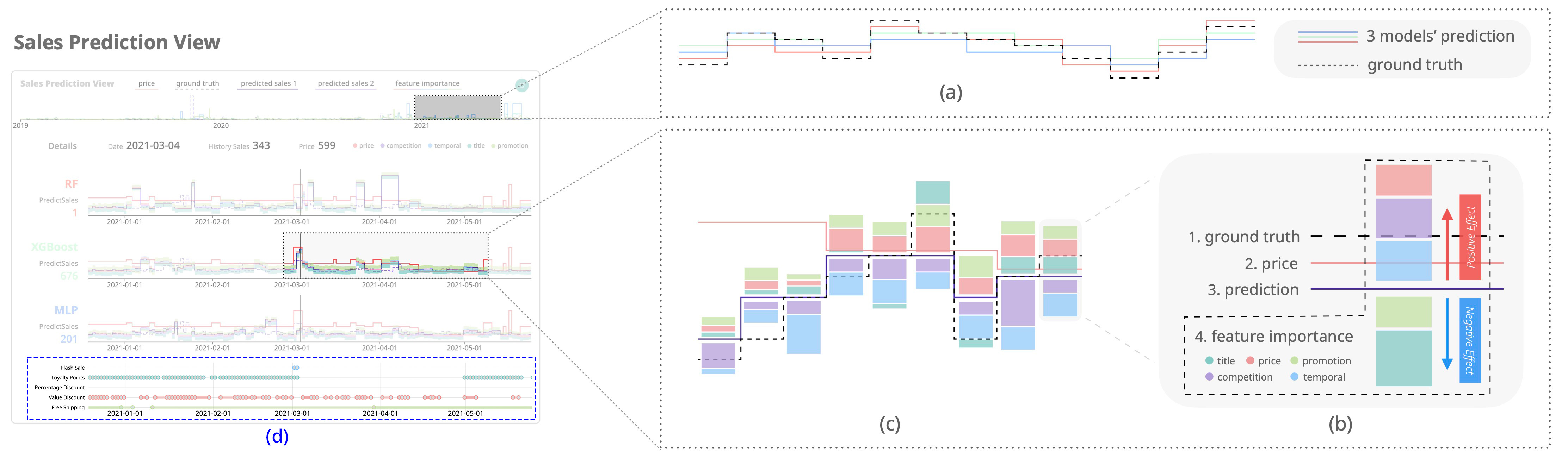}
    \vspace{-8mm}
    \caption{(a) Composition of the time horizon, (b) Visual encoding of prediction, (c) Close-up view for the sales prediction, and (d) Dotted line graph.}
    \label{fig:sales_prediction}
     \vspace{-6mm}
\end{figure*}

\subsection{Product Overview}
\par The \textit{Product Overview} (\autoref{fig:UI}(A)) enables users to efficiently select a product of interest, observe its general information and examine its relationship with other products (\textbf{R.1}). Dimensionality reduction techniques like \textit{t-SNE}, \textit{PCA}, and \textit{MDS} can generate low-dimensional representations and preserve local similarities to convey neighborhood structures like potential anomalies and clusters, all of which can be applied to explore and illustrate patterns in higher-dimensional spaces~\cite{10.1111:cgf.13417,10.1109/TVCG.2016.2598838}. Here, we use \textit{t-SNE} to project all products into two-dimensional space to see potential clusters and outlier points. After discussing with domain experts and surveying existing e-commerce practices, we used the following six metrics: \textit{sales median}, \textit{standard deviation}, \textit{quartile range}, and the \textit{correlation} between price, promotion, season and sales. These metrics form an $n$-dimensional feature vector $x_1,...,x_N$. We place the corresponding product statistics glyph, each with six features, in the calculated positions (detailed product statistics glyphs will be presented in the \textit{Competitive Analysis View} (\autoref{fig:UI}(D)). To interact with this view, the user should first set the filters that control the data, type, and brand of the product. Then, the system will automatically highlight all products that match the filter criteria for selection. The user can click on these points to select the target product. In addition, if the user knows the product ID, he/she can also enter that ID in the search box and press Search \SearchButton{}. When a specific product is selected, message tooltip such as \textit{product ID}, \textit{product name} and \textit{product category} will be displayed.

\subsection{Promotion Overview}

\par We design a \textit{Promotion Overview} (\autoref{fig:UI}(B)) to provide the analysis with the promotions of the selected product and their correlation with the sales volume in the past periods (\textbf{R.1, R.3}). As shown in \autoref{fig:promotion_overview}, there are two rings, each representing a year. The internal one represents the previous year and the external one represents the next year. The internal line graph represents the sales volume of the corresponding promotion offered on that day, and the external bar graph generates information on ``what'' and ``how many'' promotion strategies were used, with different colors representing a different types of promotions. Also, the height of the bars of the same color indicates the number of promotions using that type (\autoref{fig:promotion_overview}(a)). During the design iteration, we replace the alternative design of stacked line charts (\autoref{fig:promotion_overview}(b)), because in such a compact chart, stacked line charts can become messy. We also compare the horizontal design with the radial design. Because promotions have a cyclical dependency, and a circular glyph can better convey the cyclical pattern. In addition, the space left for daily promotions is too narrow to be clearly displayed in a horizontally stacked bar chart, which may confuse users and increase their learning costs. This radial chart allows us to compare promotion strategies and sales amounts separately along the radius for a two-year period.

\subsection{Sales Prediction View}
\par The \textit{Sales Prediction View} in \autoref{fig:UI}(C) helps users understand the detailed correlation between the five most important factors (i.e., \textit{price}, \textit{competition}, \textit{temporal variation}, \textit{title} and \textit{promotions}) and product selling amount. In particular, at the top of the view is a visualization of the time, which shows the forecast profiles for the three models (\autoref{fig:sales_prediction}(a)), making it easy for the user to adjust the length by dragging the gray slider to determine the time horizon to examine in detail. The main view of the \textit{Sales Prediction View} shows the results of the three forecast models (\textbf{R.2}) and their interpretation of sales changes. As shown in \autoref{fig:sales_prediction}(b), the graph for each model contains four parts: (1) the black dashed line (\autoref{fig:sales_prediction}(b).1) indicates the ground truth sales volume for the selected time horizon; (2) the pink solid line (\autoref{fig:sales_prediction}(b).2) shows the actual prices; (3) the purple solid step lines (\autoref{fig:sales_prediction}(b).3) represents the comparison of the predicted sales amount with the real amount; (4) five colored bars representing the importance of five characteristics are attached above and below the model prediction line (\autoref{fig:sales_prediction}(b).4) (\textbf{R.3}). It is worth noting that, as shown in \autoref{fig:sales_prediction}(b).3, the daily predicted sales amount is represented as a step line rather than a curve or line segment, so that we can use the step space as the x-axis of daily feature importance. According to \autoref{fig:sales_prediction}(c), inspired by the design of \textit{mTSeer}~\cite{xu2021mtseer}, the five feature importance are calculated, normalized, and represented as a stack of bars (\textbf{R.1}, \textbf{R.3}). Bars above the step lines indicate the positive impact on the forecast and vice versa. Further than \textit{mTSeer}, at the bottom of the \textit{Sales Prediction View} is a dotted line graph (\autoref{fig:sales_prediction}(d)), which implies the promotion time coded by the length of the dotted line. All the charts mentioned above have the same X-axis, i.e., the time axis. After selecting a specific time horizon, the user can compare the performance of different models and pick up the chart that most closely resembles the ground truth prediction in order to observe the importance of its features and the promotions on the selected days. The user can also check the newly plotted forecasts caused by the updated promotional activity settings in the \textit{Strategy Setting View} (\autoref{fig:UI}(E)).

\subsection{Strategy Setting View}
\label{sec: Strategy Setting View}
\par The \textit{Strategy Setting View} is designed to meet the user's need for promotion strategy assumptions (\textbf{R.5}). When a product is selected in the \textit{Product Overview}, all promotions within its time frame specified by the user in the sales prediction view are listed here. Users can use the category filters above to quickly locate specific promotions. For each campaign, users can press the Edit\EditButton{} button to adjust the campaign status by changing the description in the box with the formatting in \autoref{tbl:promotion_types}, or by enabling and disabling the checkboxes. Users can also modify the duration of each promotion by dragging the position and length of the slider below each term. Pressing the Delete\ \DeleteButton{} button will delete an existing campaign, while pressing the Add\ \AddButton{} button will add a campaign. When all promotions are set up, the user can 
re-run the model by pressing the Refresh\ \RefreshButton{} button and the system will draw a new step line in purple in the \textit{Sales Prediction View} and attach an updated feature importance bar to it.

\subsection{Competitive Analysis View}
\par The \textit{Competitive Analysis View} (\autoref{fig:UI}(D)) provides a comparison between the selected product and several of the most similar products in the same category (\textbf{R.1}), which can be considered as potential competitors. In this work, experts recommend comparing at least five other products of the same type. With this view, the user can find similarities and differences between the target product and its competitors, and thus build a corresponding promotion strategy solidly and confidently.

\par
\begin{figure}[h]
    \centering
      \vspace{-3mm}
    \includegraphics[width=0.45\textwidth]{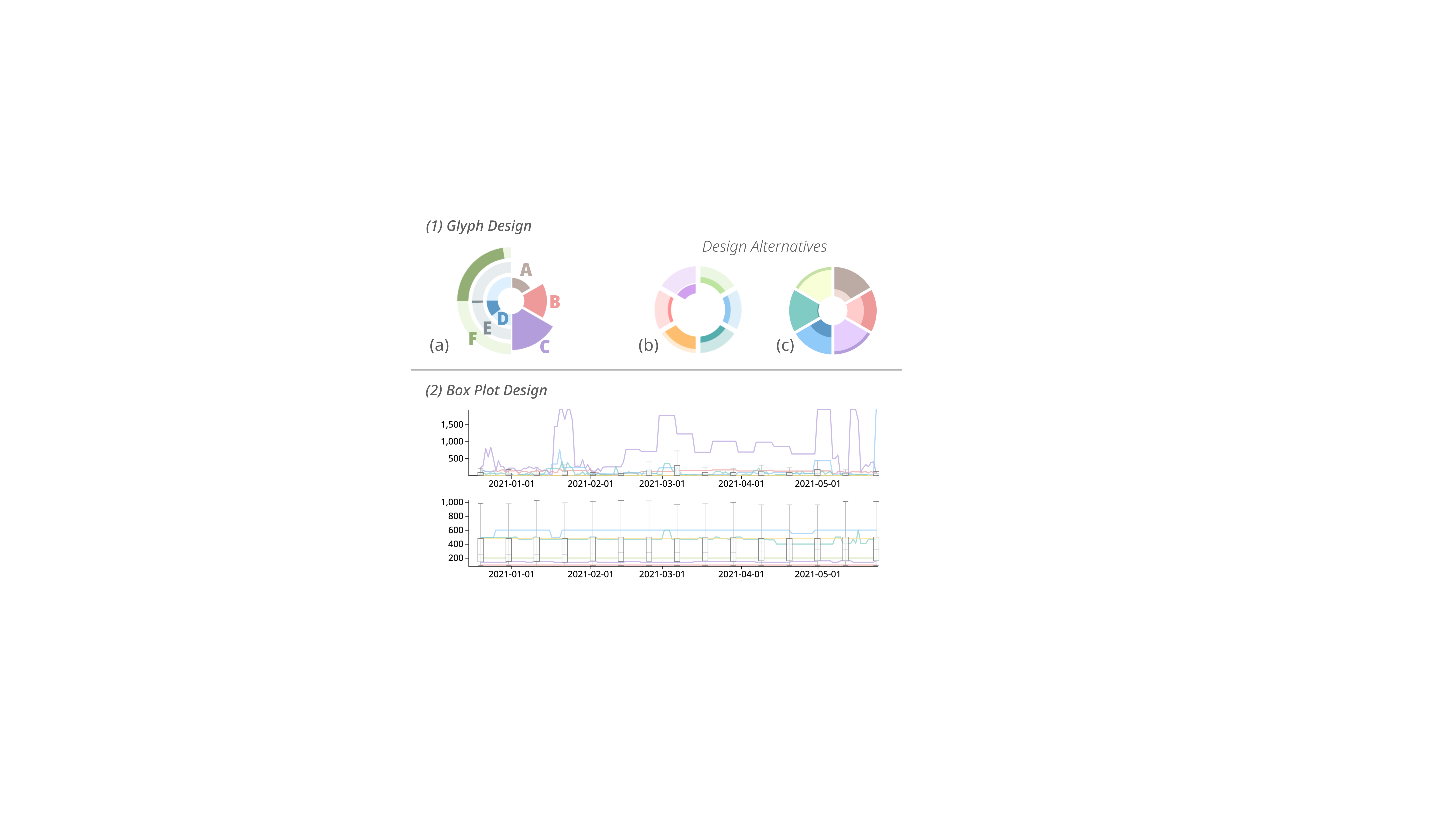}
    \vspace{-3mm}
    \caption{Competitive Analysis View: (1) Product statistics glyph design; (2) Box plot design for performance measurement.}
    \label{fig:Glyph_design}
    \vspace{-3mm}
\end{figure}

 To select the top five most similar products, we calculate the Euclidean distance between the target product and the other five products in the same category in the \textit{Product Overview} (\autoref{fig:UI}(A)). To help users get a general idea of the products and compare the differences between them, we designed a novel glyph called product statistics glyph to encode the statistical attributes of the products (e.g., stability of sales amount, price elastics). As shown in \autoref{fig:Glyph_design}(1a), the product statistics glyph is divided into two halves. For the right half, we code \textit{median} (A), \textit{standard deviation} (B), and \textit{interquartile range} (C) for the product sales amount. These three attributes are encoded as half-donut plots with the same segmentation angle because they are different features and all are positive. The height of the split donut shows the attribute value. The higher it is, the larger the corresponding value is. For the left half, we designed three donut plots representing the correlation coefficients of sales amount with price (D), promotion strength (E), and season (F), respectively. Six colors are used in the glyph to represent the six attributes of each product. Since their values range from $-1$ to $1$, we set a zero point on the horizon and define the part above the horizontal line as positive and below as negative, so that users can clearly distinguish the positive and negative effects of each of these three attributes. All values here are clamped and normalized with interquartile ranges to avoid outliers that would cause all other values to show up near zero.

\par In addition to the statistical glyphs for the products, a word cloud is constructed using all keywords for the six products, with the word size proportional to the average sales amount of all products containing the word. Two box plot-based graphs are also shown in \autoref{fig:Glyph_design}(2), where users can compare the sales amount performance and price levels of all six products individually (\textbf{R.4}). For ease of observation, we divide the time into a fixed number of intervals, i.e., ten boxes in this example, and we split the time equally on the x-axis regardless of how long the user in the sales prediction view sets the time. The total sales amount is calculated for each box on the exact date the box is positioned in the same category of the selected product. Below them is a dotted line chart, the same as the one in \autoref{fig:UI}(C), showing the promotion duration.

\par \textbf{Design alternatives.} During the iterative process of our system, we went through several design alternatives for the product statistics glyphs. Intuitively, we decided to use the glyph based on a donut chart (\autoref{fig:Glyph_design}(1)). It is divided equally into six parts, each representing a specific statistical property. Although the glyph can perfectly combine the above six data, it is difficult to set a standard for them due to the different ranges between correlations, medians, deviations, and quartiles (i.e., correlations range from $-1$ to $1$, while the range between medians, deviations, and quartiles is only positive). Initially, we drew positive bars outwards to the center circle and negative bars inwards (\autoref{fig:Glyph_design}(1b)). However, when we tried to use this glyph in the \textit{Product Overview}, the normalized bars became too small to be observed. A possible solution is to have the bars all outward and distinguish negative from positive numbers by shades of color (\autoref{fig:Glyph_design}(1c)), i.e., light shading for positive numbers and dark shading for negative numbers. However, this glyph sets up too many rules for each statistic and can be quite confusing for users to get started quickly. Finally, after discussions with domain experts, we decided to use the pie-shaped bars for purely positive values and ring-shaped bars for numbers in the $[-1, 1]$ range.

\subsection{Interactions Among the Views}
\par Rich interactions are integrated to facilitate effective in-depth analysis. (1) \textit{Zooming.} The system supports zooming in on the product overview, enabling users to locate and distinguish overlapping glyphs. (2) \textit{Filtering.} The line elements in the \textit{Sales Prediction View} can be toggled off when the corresponding legend is disabled, so that other lines can be more clearly observed. In the \textit{Product Overview}, users can filter products by category and brand to focus on specific products. (3) \textit{Brushing.} The time range in the \textit{Sales Prediction View} and the \textit{Strategy Setting View} can be adjusted interactively. (4) \textit{Parameter editing.} Users can edit promotions by modifying the text description in the input box or enabling and disabling the checkboxes in the \textit{Strategy Setting View} to perform a ``what-if'' analysis of the corresponding promotion strategy. (5) \textit{Highlighting and linking.} When the product statistics glyph in the \textit{Competitive Analysis View} is selected from the base, the system highlights the corresponding word and line graph in the box-based glyph. Thus, the user can have a clear view of the selected competitive product. (6) \textit{Hover on and tooltip.} If we hover over the model forecast graph, detailed information (such as date, sales volumes, price, etc.) is displayed. Also, if the mouse hovers over a promotion in the \textit{Sales Prediction View}, it will show the details of the promotion.

\section{Evaluation}
\subsection{Case Study}\label{sec: case study}
\subsubsection{Case I: There is No One-size-fits-all Promotion}

\par In the first case, the experts wanted to determine whether similar products always shared similar optimal promotion strategies. ``\textit{Footwear is a typical category of e-commerce marketing products, so we can start with relatively affordable shoe brands, such as 361°}'', said E2. Following this, in the \textit{Product Overview}, E2 used the drop-down boxes to filter the categories and brands to \textit{Sneakers} and \textit{361°}, and he found that one product had a higher Median and STD than the others, indicating this shoes is well-selling (\autoref{fig:case_1}(1)). Therefore, this product with ID \textit{556616451288} is then selected as the target product and its details are displayed in the information panel below(\autoref{fig:case_1}(2)). In the \textit{Promotion Overview}, E2 found that percentage discounts and value discounts were frequently applied between April and October 2020 (\autoref{fig:case_1}(3)). ``\textit{Percentage or value, which strategy is better?}'', intrigued by this, E2 adjusted the time slider in the \textit{Sales Prediction View} to dive into this period (\autoref{fig:case_1}(4)).

\par To estimate the degree of impact of the promotion activity, firstly, E2 defined the \textbf{growth rate} for a promotion: denoting the sales volume at the beginning of the promotion as ${V_t}$, and the sales volume on the previous day as ${V_{t-1}}$, the growth rate is defined as $\frac{V_t - V_{t-1}}{V_{t-1}}$, picking the sales growth on the first day as a proxy for the promotion impact.

\par In order to compare the two promotion strategies, E2 needed to ensure two things: 1) no multiple promotions are applied simultaneously; and 2) there is no overlap of the same promotion that interferes with his judgments. With dotted line graph, E2 found 5 non-overlapping zones for each promotion strategy (\autoref{fig:case_1}(5)) and calculated the average growth rate. The result shows that the growth rate of the value discount is $0.2$ and the one of the percentage discount is $0.15$, which means the value discount should be more effective for 361° sneakers.

\par To verify this assumption, E2 conducted a simulation by transferring other promotions with the same strength to the selected product to see if there was a better promotion that can further increase the sales amount. E2 picked the value discount starting from $1^{st}$ October, $2020$ (\autoref{fig:case_1}(6)), which is ``\textit{30 CNY off orders over 300 CNY}'', and changed it to ``\textit{$10\%$ off}'' (same discount rate 0.1). Then, E2 reran the prediction models (\autoref{fig:case_1}(7)). Compared with the data before the modification, he found that the prediction results of both models decreased (\textit{XGBoost} was excluded due to its high outlier rate), indicating that the value discount was more suitable for 361° sneakers (\autoref{fig:case_1}(8)).

\par To explore further, E2 turned to examine whether similar products (i.e. the product's competitors) share the same optimal promotion strategy with it. Specifically, in the \textit{Competitive Analysis View}, he explored the competitors' glyphs, associated word clouds, and box plots, respectively. The product of \textit{XTEP sneakers} with product ID \textit{40527285607} was selected because its glyph, keywords in the word cloud, and the degree of curve fluctuation in its box plot was most similar to that of our target product (\autoref{fig:case_1}(9)). E2 found that from March to June, $2020$, XTEP sneakers had a value discount promotion of $0.27$ growth rate and a percentage discount promotion of $0.35$ growth rate, which meant that the percentage discount seemed appropriate. Therefore, he used the same procedure as for the 361° sneakers (\autoref{fig:case_1}(10--11)), replacing the original percentage discount with the value discount for the XTEP sneakers, but maintaining their original strength. The prediction results of two models decreased and one remained still (\autoref{fig:case_1}(12)), indicating that the percentage discount is more suitable for XTEP sneakers (\autoref{fig:case_1}(13)). Therefore, E2 concluded that not all products have the same optimal promotion strategies, even though there are many commonalities, ``\textit{there is no one-size fits-all}''. This case also validates the ability of our system to support retailers in making predictions and comparisons about promotions.

\begin{figure*}[h]
    \centering
    \includegraphics[width=\textwidth]{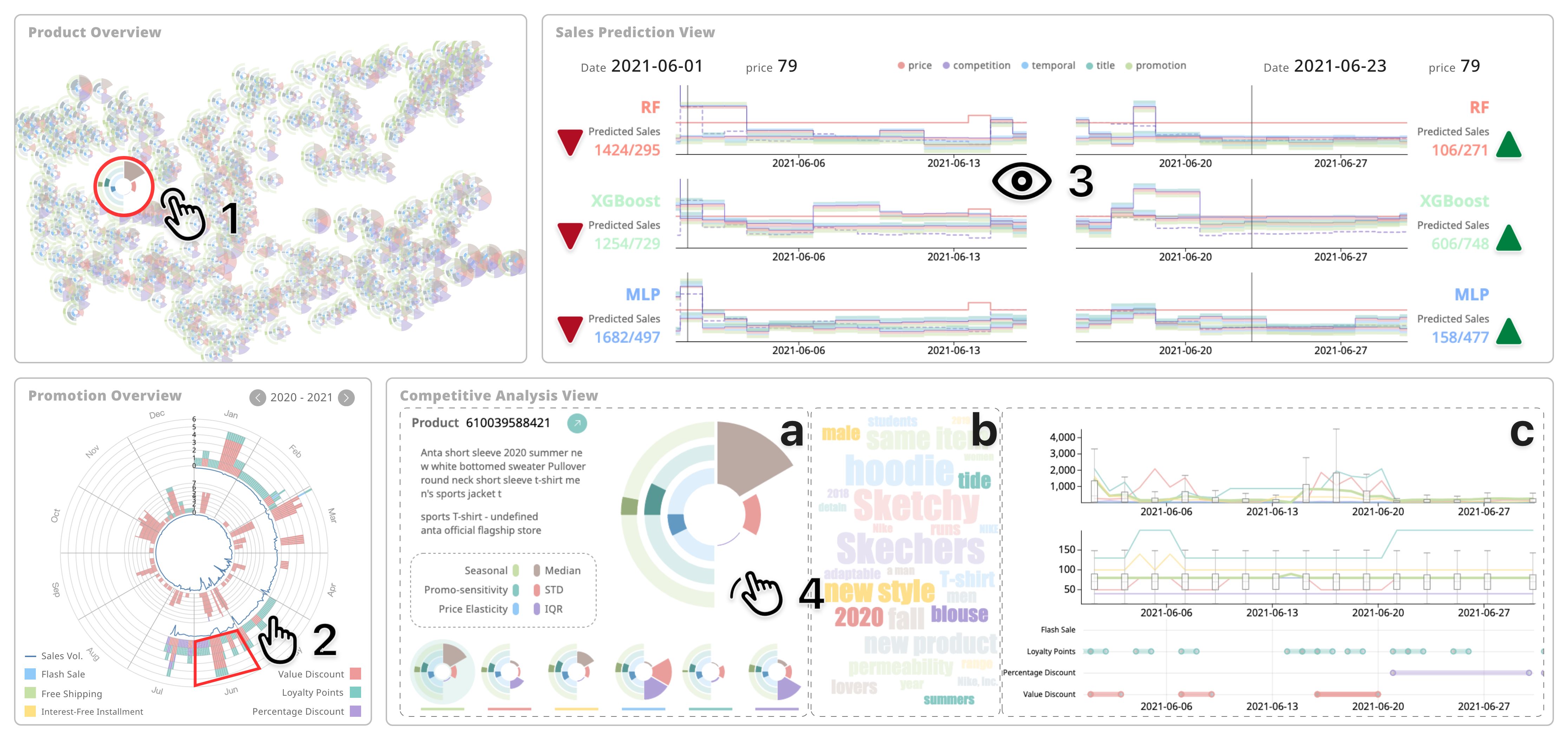} 
     \vspace{-9mm}
    \caption{Case II: (1) Select a product that has a weak correlation with the season but a strong correlation with the promotion. (2) Observe and examine the data for July. (3) Promotion before and after the conversion. After the conversion, the forecasting sales decreased at the beginning of June and showed an increase at the end of June. (4) It shows low seasonal relevance but high price and promotion relevance for this product.}
    \label{fig:case_2}
     \vspace{-6mm}
\end{figure*}

\subsubsection{Case II: Making Rational Long-Term Promotion Strategy}
\par While exploring the dataset, E1 noticed an interesting pattern in the \textit{Promotion Overview}: for some products, the promotion strategy of doing a value discount at the beginning of the promotional season and a percentage discount at the end usually performs better than applying both promotions the other way around, which aroused her enthusiasm to find out why. E1 first identified a cluster that was weakly correlated with the season but strongly correlated with the promotion in the \textit{Product Overview}. Therefore, she selected product \textit{610039588421} in the cluster since its Median (brown) sector is significantly higher than others (\autoref{fig:case_2}(1)). Then, in the \textit{Promotion Overview}, she found there are a value discount at the beginning of June $2021$ and a percentage discount at the end of the month, while the interference of loyalty points is excluded since this promotion activity continues all month long (\autoref{fig:case_2}(2)). ``\textit{This is exactly the target pattern, value discount at the beginning and percentage discount in the end}'', E1 then adjusted the time slider in the \textit{Sales Prediction View} to cover June, $2021$.

\par Using the \textit{Strategy Setting View}, E1 switched promotions so that they appear in opposite periods (i.e., shifted the value discount to the end of this month and the percentage discount to the beginning) and recorded the forecastings before and after the switch. She found that the switch leads to a decrease in the amount of sales at the beginning of the month but an increase at the end of the month compared to the chart before the switch (\autoref{fig:case_2}(3)). There are both advantages and disadvantages to each of the two strategies, so ``\textit{why do merchants tend to use the strategy value discount at the beginning and percentage discount in the end?}'' E1 implied that this strategy could be the best way for the corresponding retailers through their business experience. E2 confirmed that this commonly used strategy is related to the e-commerce schedule that seeks high sales volume at the beginning of the promotional season and requires stable sales volume at the end, because the merchants must consider the cost of storage carefully.

\par ``\textit{Are there other factors that may favor this long-term promotion strategy?}'' By looking at the glyphs in the \textit{Competitive Analysis View}, E1 confirmed that the product is neither highly seasonal nor price elastic, but relatively more sensitive to changes in promotions (\autoref{fig:case_2}(4a)). She also looked at its word cloud and found that it had very few keywords and most of them had little impact on sales, which helped the expert rule out the advertising factors with sales (\autoref{fig:case_2}(4b)). In the box plots, the sales trend for this product and its competitors in June $2021$ were very similar, which indicated that this product could be considered as a representative for the overall performance of the same category. In addition, the price of this product is relatively low (i.e., RMB $100$) and does not fluctuate significantly, which eliminates the interference of price changes on sales (\autoref{fig:case_2}(4c)). ``\textit{Most people receive their salaries at the beginning of the month, which is the time they feel most confident about spending money.}'' With more free money at their disposal, people are more likely to make impulse purchases. Therefore, doing a value discount at the beginning of the promotional season and a percentage discount at the end should be the best option for this product, ``\textit{value discounts require consumers to spend more money to meet the conditions of the offer, so it is easier to make people spend emotionally; in contrast, percentage discounts are a more rational choice}''. In addition, this interactive exploration process evoked her interest, ``\textit{while more cases are needed to summarize the hypothesis, PromotionLens has pointed the way,}'' commented E1. All in all, this case shows us that \textit{PromotionLens} can not only help users make relatively optimal promotion strategies during the long promotion season, but also capture specific patterns that ML models may fail to extract on their own.

\begin{figure*}[h]
    \centering
    \includegraphics[width=\textwidth]{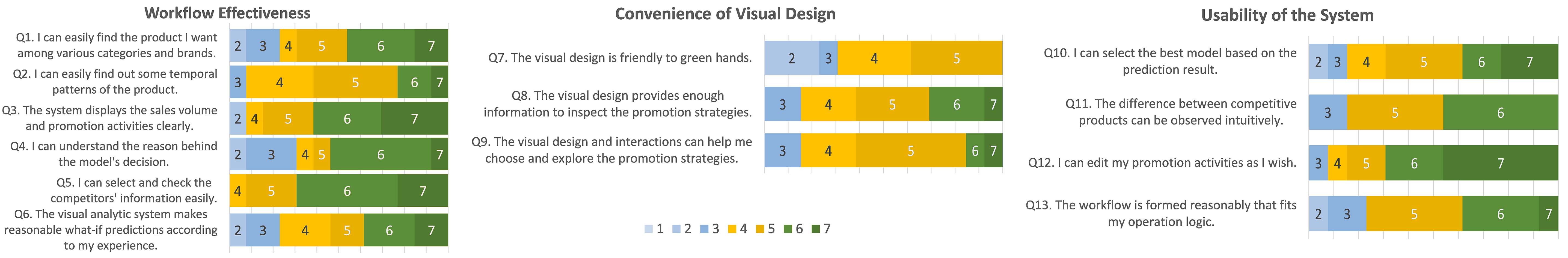}
    \vspace{-8mm}
    \caption{Results of user perceptions of \textit{PromotionLens}. $1-7$ represents ``strongly disagree'' to ``strongly agree'' for each statement.}
    \label{fig:user_study}
    \vspace{-6mm}
\end{figure*}

\subsection{User Study}
\par We conducted a user study to evaluate \textit{PromotionLens} in three aspects: \textit{workflow effectiveness}, \textit{ease of visual design} and \textit{system usability}.

\par \textbf{Participants}. We recruited $12$ participants ($4$ female, $8$ male, $age_{mean} = 23.46$, $age_{sd} = 2.93$) through word-of-mouth, consists of students and faculties from the majors of Computer Science, Economics, and Design in an university. They were all loyal online shoppers, each with approximately RMB $2000$ in annual expenditures. We selected participants with a strong interest in online shopping because they could provide us with understandable insights and help verify system usability. The study was conducted face-to-face, with users participating in tutorials, experiencing the system, performing tasks, and filling out questionnaires. Upon completion, each participant receives a $\$10$ voucher for the online e-commerce platform.  

\par \textbf{Tasks.} In our user study, each participant was asked to complete two tasks in turn. The purpose of \textbf{task 1} was to examine whether the system was capable of helping users understand the model. In this task, users were asked to find the period in which a promotion strategy had the greatest impact on sales volume and then explore the impact of each promotional activity. The goal of \textbf{task 2} is for users to develop a new promotion strategy. Participants can customize promotion strategies and check their influence on predicted sales. As our participants are not e-commerce experts or practitioners, we mainly examine the integrity and usability of the system design. For more discussion on system effectiveness, please see section \autoref{sec: case study} and \autoref{sec: expert interview}.

\par \textbf{Procedure.} The whole process lasted about one hour. Before we presented our system, all participants were informed that their reactions and feedback about our system would be collected anonymously in a questionnaire. Then, we conducted a tutorial session for about $20$ minutes, during which we introduced the workflow of our system, explained all the operations for each view, and told the users which types of features were worthy of attention. We then allowed participants to pre-operate the system and ask us further questions for around $10$ minutes. When the participants felt ready, they were given the two tasks mentioned above, which they needed to complete individually. After completing these two tasks, we would distribute a questionnaire with questions using Likert scale criteria~\cite{OBRIEN201828}, with $1$ representing ``strongly disagree'' and $7$ representing ``strongly agree'' (\autoref{fig:user_study}). The items measured in the questionnaire included $6$ items on the effectiveness of our system workflow, $3$ items on visual design convenience, and $4$ items on system usability~\cite{xia2019peerlens,sun2020dfseer}. Other comments that did not participate in the questionnaire were also recorded for reference.

\par \textbf{Results.} The questionnaire results are presented in \autoref{fig:user_study}. In general, participants rated \textit{PromotionLens} highly. In terms of workflow effectiveness, participants felt that \textit{PromotionLens} could display sales volume and promotions (Q3) and that they could easily select and view information about competitors (Q5). In addition, the visual design of the system generally facilitated the provision of sufficient information to examine promotion strategies (Q8) and aided exploration and decision making (Q9). However, some were concerned that the visual design was not very green hand friendly (Q7), with the main concerns being that ``\textit{information is a bit overwhelming for first-time users, but should not be too difficult after getting familiar with the system}''. Regarding system usability, participants found it quite useful to be able to select the best model based on the predictions (Q10), to observe the differences between competing products (Q11), to edit promotions as desired (Q12), and to reasonably form a workflow (Q13).

\subsection{Expert Interview} \label{sec: expert interview}
\par We conducted semi-structured interviews with E1 -- E5 to check whether \textit{PromotionLens} helped explore the impact of promotion strategies.

\par \textbf{System Performance.} All experts agreed that the system runs smoothly and the visualization is decent and organized. ``\textit{The system is useful for the seller's promotion strategy development}''. Our approach proposes an innovative ``what-if'' analysis, which is helpful for online retailers like E4 and E5. In discussing with experts what most enlightened them about our system, E1 said that extracting ``unexpected'' factors such as seasonality, which data engineers may not normally notice, may reveal new associations, allowing researchers like her to validate existing associations or propose new marketing theories.

\par \textbf{Learning Curve.} The average time spent presenting our system to domain experts was about $15$ minutes. After the presentation, we gave the experts $20$ minutes to freely explore each view in \textit{PromotionLens}, and they all found the system easy to work with. ``\textit{The information provided by \textit{PromotionLens} was sufficiently reasonable for most of our analysis work, and our previous suggestions and requests were taken into account}''. The experts also said that with a basic tutorial, most users would get started on their own with their customized learning habits. During the discovery stage, E2, working as a product manager (PM), needed to contact the management team to raise concerns that some users may be intimidated and frightened by the unfamiliar workflow. ``\textit{However, if they are given a tutorial session to show them the functionality of each view, they would soon become experts.}'' For further iterations, he suggested that we implement a report generation feature to summarize the identified patterns and outliers so that users without much technical background can get results quickly and efficiently.

\par \textbf{Generalizability and Scalability.} In the interviews, we asked about which components in \textit{PromotionLens} could be directly transferred to other scenarios; three aspects were considered valuable: (1) \textit{Visualization}. E1 commented that the designs in the \textit{Product Overview} and the box plot used in \textit{PromotionLens} are also widely used in other visualization systems, so they should be easily generalized to other similar scenarios. (2) \textit{Data}. E3 noted that currently, our system only works on $500$ products, but other product data can be easily inserted because they share the same format. Therefore, our system has a strong potential to include multiple products. (3) \textit{Algorithms}. E3 found it helpful that \textit{PromotionLens} has an explicit procedure to handle the dataset and the system is able to switch or add new models when necessary. In addition, E1 suggested that our system could be extended to explore additional features beyond promotion strategies, e.g., exploring the relationship between seasonal purchase and product sales volume. Since we have only entered the top $500$ best-selling products in the system, experts noted that scalability could be an issue for the \textit{Product Overview} and that ``if the dataset contains thousands of products, the view should be very crowded''. We can hide the filtered products instead of highlighting the selected products. If the view is still crowded, we can use multiple hierarchies to reduce the visual clutter.

\section{Discussion and Limitation}
\par \textit{PromotionLens} considers most of the explicit factors that can influence sales volume; it combines time, product name, competitors, and historical sales volume with promotion strategies and gives a feature importance analysis along with the forecast. However, more implicit factors might also lead to fluctuations in sales volume. For example, social factors (e.g., news events, macroeconomics) may affect the sales volume of certain categories of products. In addition, according to Forbes, on-site promotional sales are becoming increasingly popular in online shopping~\cite{naeem2021social}, thus making it more important for e-commerce to weigh in when considering promotion strategies. Researchers can attach more implicit factors to our work for further study.

\par \textbf{Intention of Promotions.} Usually, there is only one criterion to evaluate a promotion strategy: whether it significantly increases sales. However, every product has its own life cycle. promotion strategies should be designed to serve the trends at each stage. A roller-coaster sales curve is the last thing a retailer wants, because rapid rises and falls can put tremendous pressure on warehousing, which will exponentially increase the cost of the product. Similarly, rapid sales growth may leave products out of stock, and producing another batch would cost extra and increase the risk of overstocking. Therefore, rational sellers tend to have a ``hat-like'' sales curve~\cite{10.2307/1880689} (i.e., \textit{introduction}, \textit{growth}, \textit{maturity}, and \textit{decline} phases) and maximize profits. This lifecycle inspires future iterations where we can implement another input panel that allows \textit{PromotionLens} to accept data on e.g., manufacturing cost, storage cost, to help retailers find the best promotion strategies.

\par \textbf{Inspirations for Marketing.} The emergence of monthly promotion model inspired our domain experts to hypothesize that factors such as buyers' financial status and buyers' spending habits may influence the effectiveness of promotion strategies and further affect sales volume. Thus, \textit{PromotionLens} is able to reveal hidden relationships and interesting phenomena for further academic research in these fields.

\par \textbf{Limitation.} There are several limitations to this work. First, our system only works for the top $500$ best-selling sports products from a few well-known brands. However, in addition to these brands, there are many mid-sized or small stores on this platform that need help with promotional production. Large e-commerce stores have a large volume of sales every day, while sales data for smaller retailers can be very sparse and our system needs further testing. Second, when quantifying promotions, the system does not take into account the limitations between promotions, such as not being able to use value discounts when users choose percentage discounts. A more comprehensive quantification method needs to be applied to our system in the future. Third, we recruit participants (mainly students) who lack expertise to do the user study, resulting in limited validity of the quantitative evaluation of ``the effectiveness of the workflow''. Fourth, for simplicity, we narrow the scope of the study to footwear and apparel, and as the amount of data expands, we can study different marketing models in more areas.

\section{Conclusion and Future Work}
\par In this study, we present \textit{PromotionLens} to support retailers and researchers in exploring, comparing, and modeling their promotion strategies. It combines representative multivariant time-series forecasting models and visualizations to demonstrate and explain the impact of sales and promotional factors, and also supports ``what-if'' analysis of promotions that can inspire marketing researchers and help online retailers. In the future, we will access data on manufacturing cost and storage cost to maximize profits for users. We will also integrate other heterogeneous data, such as news events and macroeconomics influences, to reveal causes more accurately.

\acknowledgments{
We are grateful for the valuable feedback and comments provided by the anonymous reviewers. This work is partially supported by the Research Start-up Fund of ShanghaiTech University, the Fundamental Research Funds for the Central Universities, Sun Yat-sen University under Grant No.: 22qntd1101, and HKUST-WeBank Joint Laboratory Project Grant No.: WEB19EG01-d.}

\bibliographystyle{abbrv-doi}

\balance
\bibliography{template}
\end{document}